\documentclass[prl,aps,epsfig,twocolumn,showpacs]{revtex4}
\usepackage{graphicx}
\usepackage{amssymb}
\usepackage{amsmath}
\usepackage{ulem}
\usepackage{color}

\begin{document}

\draft

\title{One-dimentional magnonic crystal as a medium with magnetically tunable
disorder on a periodical lattice}

\author{J. Ding$^1$, M. Kostylev$^2$, and A. O. Adeyeye$^1$}
\affiliation{$^1$Information Storage Materials Laboratory,
Department of Electrical and Computer Engineering, National
University of Singapore, Singapore 117576}

\affiliation{$^2$School of Physics, University of Western Australia,
Crawley, Western Australia 6009, Australia}

\date{\today}

\begin{abstract}

\indent We show that periodic magnetic nanostructures (magnonic
crystals) represent an ideal system for studying excitations on
disordered periodical lattices because of the possibility of
controlled variation of the degree of disorder by varying the
applied magnetic field. Ferromagnetic resonance (FMR) data collected
inside minor hysteresis loops for a periodic array of Permalloy
nanowires of alternating width and magnetic force microscopy images
of the array taken after running each of these loops were used to
establish convincing evidence that there is a strong correlation
between the type of FMR response and the degree of disorder of the
magnetic ground state. We found two types of dynamic responses:
anti-ferromagnetic (AFM) and ferromagnetic (FM), which represent
collective spin wave modes or collective magnonic states. Depending
on the history of sample magnetization either AFM or FM state is
either the fundamental FMR mode or represents a state of a magnetic
defect on the artificial crystal. A fundamental state can be
transformed into a defect one and vice versa by controlled
magnetization of the sample.\

\end{abstract}

\vspace{5 cm}


\maketitle

\indent Arrays of magnetic nanowires (NW) have generated
considerable scientific interest due to their potential application
for microwave devices ~\cite{Choi2007,Zhang2009} and domain wall
logic ~\cite{Zhu2005}. On the other hand they represent a variety of
artificial crystals \cite{Yablonovich} in which wave excitations can
propagate. Magnetic artificial crystals are called magnonic
crystals(MC), the wave excitations of which are collective spin wave
(magnonic) modes which exist in the microwave frequency range
\cite{Nikitov,review, Neusser2009, Topp2010, Kruglyak}. In contrast
to photonic \cite{Yablonovich} and sonic \cite{Miyashita} crystals
magnonic crystals are easily frequency tunable by applying magnetic
field to the structure. MC response at remanence currently attracts
a lot of attention (\cite{Tacchi2010, Topp2010, Zivieri, Otani}).
First, minimizing the bias magnetic field applied to a nanostructure
is important for possible applications in tunable microwave devices.
Second, magnetic periodic nanostructures may have a number of
periodic magnetic configurations (magnetic ground states) for the
same periodic geometry of the material, and the material can be
switched between these ground states. Controlling of the magnonic
frequency gap by switching between two ground states \cite{Topp2010}
and reconfiguration of a magnonic crystals has been recently
demonstrated (\cite{Tacchi2010, Zivieri}). Importantly, in these
previous experimental studies additional microwave responses were
seen within the hysteresis loops which are not present in the
theories which assume a perfect magnetic  periodic order for the
array \cite{Tacchi2010, Topp2010}. It was supposed that these
branches originate from \textit{magnetic} disorder on the periodic
lattice. In the present paper we will experimentally show that these
branches indeed originate from deviation from a perfect magnetic
periodic order. This deviation takes the form of "magnetic defects"
which represent individual wires or small clusters of wires in which
magnetization vector points in a wrong direction with respect to the
average magnetic order of MC.

Defects on a crystal lattice are objects
of fundamental importance: this can result on week localization of
electron on a crystal lattice in a condensed matter \cite{Niimi};
Anderson localization can occur due to disorder in different
systems, such as condensed matter \cite{Anderson}, Bose-Einstein
Condensate \cite{billy}, and photonic crystals \cite{John, Lahini}
(see also extensive literature in all those papers). Importantly, to
study the effects of disorder and defects on a crystal lattice is
not simple. Often one has to fabricate a large number of samples
with different degrees of disorder, but otherwise identical, like it
was recently done in Ref.\cite{Niimi}.

\begin{figure}[tb]
\setlength{\abovecaptionskip}{0pt}
\setlength{\belowcaptionskip}{0pt}
\begin{center} \resizebox*{8.5cm}{!}{\includegraphics*{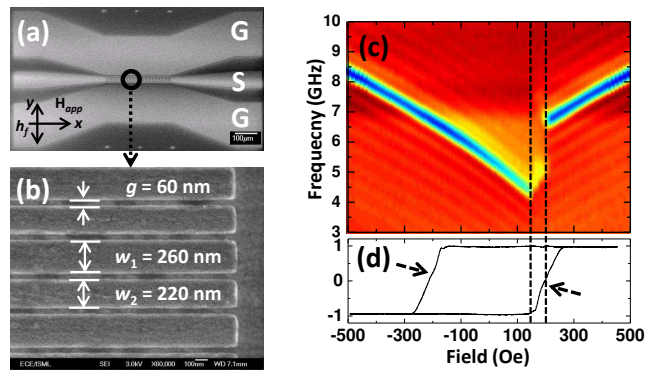}}
\caption[]{(Color Online): (a): SEM image of the CPW showing the
ground - signal - ground (G-S-G) lines. (b): SEM image of the
alternating widths nanowires array (\textsl{w$_{1}$} = 260 nm,
\textsl{w$_{2}$} = 220 nm and edge-to-edge separation \textsl{g} =
60 nm). (c): Full loop 2D FMR absorption spectra for the NW arrays.
(d): Normalized M-H loop for the array.} \label{Fig1}
\end{center}
\end{figure}
\indent In this paper we investigate the effect of magnetic disorder on
magnonic excitations on the lattice of a 1D MC in the form of a
dipole coupled array of magnetic nanowires \cite{review}. We
demonstrate that this material represents an excellent model system
for studying disorder because of the possibility of controlled
variation of disorder degree by varying a magnetic field applied to
the sample.

\indent Array of densely packed Ni$_{80}$Fe$_{20}$ NWs was
fabricated directly on top of a coplanar waveguide (CPW) using
electron beam lithography, E-beam evaporation and lift off
processing. Figs.1(a-b) shows the SEM images of the CPW and two
periods of alternating width NWs. The X, Y and Z (out-of-plane) axes
of a Cartesian frame of reference are parallel to the length
\textsl{l}=10 $\mu$m, width \textsl{w} and thickness \textsl{t} = 30
nm of NWs, respectively. Each period consists of two NWs with
different widths \textsl{w$_{1}$} = 260 nm and \textsl{w$_{2}$} =
220 nm. The edge-to-edge spacing \textsl{w$_{g}$} is 60 nm. Thus,
the structure period is \textsl{w$_{a}$} =
\textsl{w$_{1}$}+\textsl{w$_{2}$}+2\textsl{w$_{g}$} = 600 nm. The
array size in the direction of the array periodicity is 20 micron.
The magnetic ground state of NW arrays was characterized using a
magneto-optical Kerr effect magnetometer (MOKE) with a laser spot
size of about 10 $\mu$m and by magnetic force microscopy (MFM). The
FMR responses was measured in the 1 - 20 GHz range using a microwave
vector network analyzer (VNA) Agilent 8363C. To obtain a high
frequency response, VNA is connected to CPW by a
ground-signal-ground (GSG)-type microwave probe similar to that
reported in Ref. ~\cite{Topp2010}. The magnetic radio-frequency
field \textsl{h$_{f}$} of CPW is applied along the Y direction,
while an external static magnetic field (\textsl{H$_{app}$}) is
along the X direction as shown in Fig.1(a). FMR measurements were
performed at room temperature by sweeping the frequency for fixed
\textsl{H$_{app}$}. This was repeated for a number of minor
hysteresis loops for the sample. Each time we started from the
negative saturation field -\textsl{H$_{sat}$} = -500 Oe, passed
through zero, and then gradually increased the field to a maximum
\textsl{H$_{max}$} $<$ \textsl{H$_{sat}$} (forward half of a loop).
The field is then subsequently decreased to -\textsl{H$_{sat}$}
(backward half of a loop). A number of minor loops with different
values of \textsl{H$_{max}$} were run. At the end of each loop an
MFM image of the remanent state was also taken after the
\textsl{H$_{app}$}  had been increased again from
-\textsl{H$_{sat}$} to the same \textsl{H$_{max}$} and then switched
off to achieve the remnant state.

\indent Fig.1(c) displays the absorption spectra for the major
hysteresis loop for the sample (\textsl{H$_{app}$} ranges from -500
Oe to +500 Oe which corresponds to the forward half of the loop). A
stepwise increase in the resonance frequency is seen in the panels
between 140 and 210 Oe and can be attributed to the reversal of
magnetization in the wider wires ~\cite{Tacchi2010}. A magnetic
ground state which is characterized by alignment of the static
magnetization vector with \textsl{H$_{app}$} for the wider stripes
and the counter-alignment of them for the narrower stripes
("Antiferromagnetic" (AFM) ground state) is expected for this field
range. The normalized MOKE data shown in Fig.1(d) are in agreement
with this type of the ground state. From this panel one sees that
the slope of the hysteresis loop slightly changes around
\textsl{H$_{app}$} = 170 Oe (indicated by the dashed arrows) which
evidences transition to the AFM state. Importantly, we do not see
formation of a plateau, which was observed in the hysteresis loops
in Ref. ~\cite{Goolaup2007} (Figs.3 and
4 in that paper). This suggests the AFM state is not stabilized for
a range of applied fields in the present case of small values for
$\Delta$\textsl{w} and \textsl{g}. Rather, in agreement with a small
squareness of the hysteresis loop in Fig.1(d) a gradual transition
from a ferromagnetic (FM) ground state through an AFM one to a new
FM state is expected.

\begin{figure}[tb]
\setlength{\abovecaptionskip}{0pt}
\setlength{\belowcaptionskip}{0pt}
\begin{center} \resizebox*{8.5cm}{!}{\includegraphics*{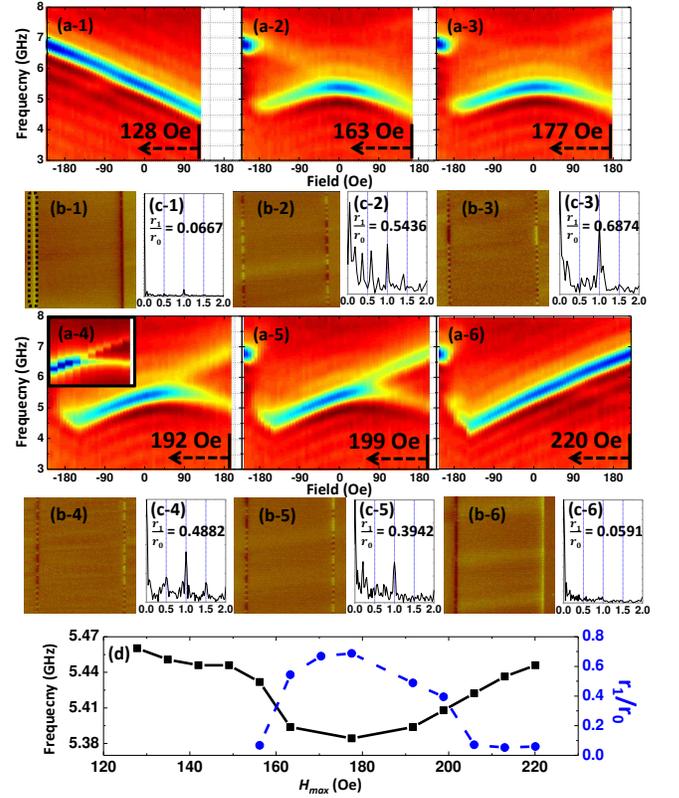}}
\caption[]{(Color Online): (a-1) to (a-6): FMR absorption spectra
inside the minor loops with \textsl{H$_{max}$} = 128 Oe, 163 Oe, 177
Oe, 192 Oe, 199 Oe and 220 Oe. (b-1) to (b-6): MFM images for the
same \textsl{H$_{max}$} at remanence. (c-1) to (c-6): Fourier
transforms of the respective MFM data.(d) Frequency of the
fundamental mode at remanence (solid line) and the ratio
r$_{0}$/r$_{1}$ (dashed line) as a function of \textsl{H$_{max}$}. }
\label{Fig2}
\end{center}
\end{figure}
\indent In Figs.2(a-1) to (a-6), we show the FMR absorption spectra
within the minor loops (\textsl{H$_{app}$} ranges from -500 Oe to
\textsl{H$_{max}$}, with 128 Oe $\leq$  \textsl{H$_{max}$} $\leq$
220 Oe for different loops). For clarity, only the backward half of
the loop, for which \textsl{H$_{app}$} decreases from
\textsl{H$_{max}$} to -\textsl{H$_{sat}$}, is shown. The forward
halves of the loops are trivial: each of them coincides with the
section \textsl{H$_{app}$} $\leq$  \textsl{H$_{max}$} of the
spectrum in Fig.1(c) for the respective value of \textsl{H$_{max}$}.

\indent Panels b-1 to b-6 of Fig.2 display the corresponding MFM
images taken at remanence. The wires are oriented horizontally. Two
vertical lines close to the image edges with a contrast different
from the rest of the area ("edge contrast lines") are signatures of
the wire edges. A bright spot is consistent with the static
magnetization vector pointing away from the spot and the dark spot
is for the tip of magnetization vector pointing to the respective
wire edge. First one observes that for each bright/dark spot at one
edge of the array there is a respective spot of the opposite
contrast (dark/bright) at the opposite edge. This fact together with
an absence of signatures of stray fields of domain walls inside the
array evidences that all the wires are in a monodomain state in all
panels.

\indent The degree of magnetic disorder varies between the panels.
We quantify the disorder degree by performing a Fourier transform of
the distribution of the MFM contrast along one of the edge contrast
lines. Panels c-1 to c-6 display the Fourier spectra along with the
ratio of the amplitudes $r_1/r_0$ of two important Fourier
harmonics. One ($r_1$) corresponds to the Fourier wavenumber $k_1$
equal to the lattice vector G=$2 \pi/(w_1+w_2+2g)$, and the other
($r_0$) to the zero wavenumber $k=0$. One sees that the richest
spectrum displaying a largest number of harmonics  for $k/G<1$ is in
Panel c-2. Indeed, the edge contrast for this state is the most
irregular. Additional harmonics but of smaller relative amplitudes
than in panel c-2 are also seen in Panels c-4 and c-5. They
correspond to $k=G/2, k=G/3$, and $G/4$. By comparing these data
with the respective MFM images one finds that the ground states are
more regular in these cases. Panel c-3 displays just two prominent
peaks for $k=0$ and for $k=G$ and is characterized by the maximum
value of $r_1/r_0$ (see Fig.2(d)). The Fourier peak at $k/G=1$
corresponds to the AFM order. Indeed, the respective MFM image
displays an almost perfect AFM order.

\indent Let us first concentrate on the FMR response in Panels
2(a-1) and 2(a-6). Panel 2(a-1) is characterized by a single
monotonically falling branch ($\omega(H_{app})$). The respective MFM
image shows that this response originates from a ground state with a
perfect ferromagnetic order with the static magnetization vector
pointing to the right (Fig.2(b-1)). Similarly, the growing branch in
Fig.2(a-6) existing for of $\omega(H_{app})$ for \textsl{H$_{app}$}
\textgreater -120Oe  is identified as a FMR response of the
perfectly ferromagnetic ground state with the static
magnetization vector points to the left (Fig.2(b-6)). This
identification is in agreement with the theory of collective modes
for the FM state \cite{Topp2010, Tacchi2010}.

\indent Let us now discuss Panels 2(a-3) and 2(b-3). The
(\textsl{H$_{app}$}) curve for this minor loop is non-monotonic. In
Ref.~\cite{Topp2010} this branch has been identified as the
fundamental acoustic collective mode for the anti-ferromagnetic
state. From Fig.2(b-3) one sees that, indeed, the ground state for
this minor hysteresis loop is antiferromagnetic. This represents an
experimental evidence for the previous identification
\cite{Topp2010} which was  based solely on the theory. The
experimentally observed state is actually not perfectly AFM, a
3-period long ferromagnetic defect is observed in the middle of the
MFM image. This suggests that a faint response existing for
$H_{app}>0$ and monotonically growing in frequency with $H_{app}$
may be the dynamic response of the FM defect.

\begin{figure}[tb]
\setlength{\abovecaptionskip}{0pt}
\setlength{\belowcaptionskip}{0pt}
\begin{center} \resizebox*{8cm}{!}{\includegraphics*{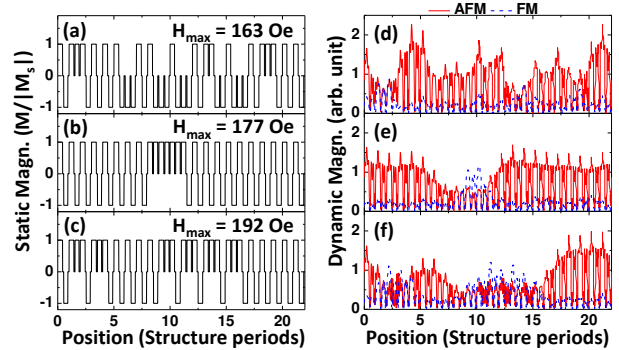}}
\caption[]{(Color Online): (a-c): The magnetic ground states for
\textsl{H$_{max}$} = 163 Oe, 177 Oe and 192 Oe, respectively. (d-f):
the respective calculated profiles of dynamic magnetization. Red
solid line: AFM mode; blue  dashed line: FM mode.} \label{Fig3}
\end{center}
\end{figure}
\indent Later we will confirm this conclusion by a numerical
simulation. But first let us discuss the responses for the states
which are transitional between the well ordered FM and the AFM
states (Figs.2(a-2), 2(a-4) and 2(a-5)). The main observations which
can be drawn from these panels are as follows. (i) The FMR response
for the transitional states looks like combination of the responses
for the FM and the AFM states in Figs.2(a-1), 2(a-6) and 2(a-3),
such that the respective branches can be termed "FM" and "AFM"
branches. (ii) Linewidths for the observed resonance peaks do not
vary noticeably with \textsl{H$_{app}$} across each minor loop and
between the loops and are very close to the linewidth obtained on a
reference continuous film (the full linewidth is 0.5GHz). (iii) The
shape of the AFM branch and the mode frequency at remanence varies
between the loops (Fig.2(d)) and is correlated to $r_1/r_0$.

\indent In order to explain these features numerical simulations
based on a 1D version of the model from ~\cite{Tacchi2010-1} have
been performed. The ground states for the linear dynamics
simulations were assumed to be the same as observed in the
respective MFM images (Figs.2(b-1) to (b-6)). To this end the
experimental contrast line data were digitized such that +1
corresponds to static magnetization vector pointing to the right and
-1 to one pointing to the left (Fig.3(a-c)). Periodic boundary
conditions (PBC) at both ends of the digitized contrast lines were
assumed. In the dynamic calculations using this ground state model
we utilized the values for the saturation magnetization and for the
resonance linewidths extracted from the FMR data obtained on a
non-nanostructured reference sample.

\indent An example of the results of simulation of a frequency vs.
applied field dependencies is shown as an inset to Fig.2(a-4). One
sees a good qualitative agreement with the experiment. The simulated
mode profiles for the disordered state are complicated (Fig.3). The
simplest ones are observed for the ground state of Fig.2(b-3). As
seen from Fig.3(b), the AFM ground state contains an FM defect which
is 3 structure periods long. Accordingly, one observes one extended
dynamic state of AFM nature (Fig.3(e)) whose amplitude noticeably
decreases on the FM defect. This mode is responsible for the main
peak of the FMR response. The defect has its own mode. Localization
of the defect mode on the defect is relatively strong: the amplitude
of the mode half a way between two neighboring defects (recall PBC)
is about 20$\%$ of the maximum inside the defect. For this reason
the FMR response of this mode is weak, and is consistent with a
faint FM-type response seen in Fig.2(a-3). It is appropriate to term
this weak response as an "impurity" or "defect" state of MC.

\indent The ground state in Fig.3(a) (corresponds to Fig.2(b-2)) can
be characterized as an AFM one with multiple single-site FM defects.
From Fig.3(d) one sees that the dominating mode is the
lower-frequency AFM mode. The higher-frequency FM modes has
considerably smaller excitation amplitude which is in full agreement
with Fig.2(a-2). Both modes are more extended than in Fig.3(e) and the
picture shows less regularity, but still localization of particular
modes on particular clusters, FM or AFM ones, is clearly seen.
Again, one can characterize the FM mode as an impurity state based
on the above considerations.

\indent The ground state in Fig.3(c) (corresponds to Fig.2(b-4)) can
be regarded as two defectless AFM clusters (4th-9th periods and from
16th period onwards) and two FM clusters (2-4) and (9-15) each
having a number of single-site AFM defects (1 and 3 defects
respectively).  As in Fig.3(d) one observes that a small number of
defects does not prohibit formation of a fundamental collective
mode, in this case of one of the FM type. Both AF and
defect-containing FM dynamic states are again extended, with a
noticeable amplitude on the clusters of the different type. The
excitation amplitudes of these two modes are comparable which is in
agreement with the experimental data in Fig.2(a-4). This case
clearly demonstrates potential of the wire arrays as a model medium
for studying effects of disorder: a simple magnetization procedure
transforms a state of AFM type with FM impurities (Fig.3(d)) into a
state of FM type with AFM impurities.

\indent Our simulation has also shown that mode frequencies for the
transitional states noticeably deviate from ones for the respective
frequencies for the array in the perfect FM and AFM states. This is
in agreement with the experimental data in Fig.2(d), where variation
of frequency of the main peak at remanence is shown as a function of
\textsl{H$_{max}$} for the respective loop. In particular,
considering a single FM cluster of the type in Fig.3b inside an AFM
array in the simulation we found that the frequency of a mode
localized on the cluster tends to the frequency for an infinite
array with the FM order with increase of the defect length. This
suggests that the observed variation in frequency is due to an
effect of confinement of the cluster inside the array (similar to
confinement effect for a single wire \cite{Guslienko}) and of its
interaction with the AFM environment. The same applies to the
fundamental (AFM) mode: its frequency is the smallest when no defect
is present and increases due to confinement (recall PBC) as the FM
defect grows.

\indent In conclusion, we studied magnetic disorder on the arrays of
dipole coupled nanowires. We found experimental evidence that
switching of nanowires produces a disordered magnetic ground state
inside the hysteresis loop. Collective dynamic states localized on
clusters with different magnetic orders are formed and the
respective FMR response inside the hysteresis loop represents a
doublet, whereas outside the loop the fundamental oscillation is a
FM singlet. Depending on the magnetization history either AFM or FM
state of the doublet can be either a fundamental or a defect
("impurity") one. Localization of these dynamic states on the
respective clusters is not very strong, the states extend into the
clusters of different order such that there amplitude is noticeable
everywhere. Individual magnetic defects inside the clusters do not
prohibit formation of collective modes.

Acknowledgment: we acknowledge financial support from the National
Research Foundation, Singapore under Grant No. NRF-G-CRP 2007-05,
and from the Australian Research Council.


\begin{thebibliography}{25}

\bibitem{Choi2007}
S. Choi \textsl{et al.}, Phys. Rev. Lett. \textbf{98}, 087205 (2007)

\bibitem{Zhang2009}
H. Zhang \textsl{et al.}, Appl. Phys. Lett. \textbf{95}, 232503
(2009)

\bibitem{Zhu2005}
X. Zhu \textsl{et al.}, Appl. Phys. Lett. \textbf{87}, 062503 (2005)

\bibitem{Yablonovich}
E. Yablonovich \textsl{et al.}, J. Opt. Soc. Am. B, \textbf{10}, 283 (1993).

\bibitem{Miyashita}
T. Miyashita, Meas. Sci. Technol. \textbf{16}  R47(2005)

\bibitem{Nikitov}
S. A. Nikitov \textsl{et al.}, J. Magn. Magn. Mater. \textbf{236}
320 (2001).

\bibitem{review}
G. Gubbiotti \textsl{et al.}, J. Phys. D: Appl. Phys. \textbf{43}
264003 (2010).

\bibitem{Neusser2009}
S. Neusser \textsl{et al.}, Adv. Mater. \textbf{21}, 2927 (2009).


\bibitem{Kruglyak}
V.V. Kruglyak \textsl{et al.}, Phys. Rev. Lett. \textbf{104}, 027201
(2010)

\bibitem{Topp2010}
J. Topp \textsl{et al.}, Phys. Rev. Lett. \textbf{104}, 207205
(2010)

\bibitem{Tacchi2010}
S. Tacchi \textsl{et al.}, Phys. Rev. B \textbf{82}, 184408 (2010)

\bibitem{Zivieri}
R. Zivieri, F. Montoncello, L. Giovannini, et al., "Collective spin
modes in chains of dipolarly interacting rectangular magnetic dots",
in print in Phys. Rev. B.

\bibitem{Otani}
A. Barman \textsl{et al.}, J. Phys. D: Appl. Phys. \textbf{43}
422001 (2010).

\bibitem{Niimi}
Y. Niimi \textsl{et al.}, Phys. Rev. Lett. \textbf{102} 226801
(2009).

\bibitem{Anderson}
P. W. Anderson, Phys. Rev. \textbf{109} 1492 (1958).

\bibitem{billy}
J. Billy \textsl{et al.}, Nature, \textbf{453}, 81 (2008).

\bibitem{John}
S.John, Phys. Rev. Lett. \textbf{58} 2486 (1987).

\bibitem{Lahini}
Y. Lahini \textsl{et al.}, Phys. Rev. Lett. \textbf{100}, 013906
(2008).

\bibitem{Goolaup2007}
S. Goolaup \textsl{et al.}, Phys. Rev. B \textbf{75}, 144430 (2007).

\bibitem{Tacchi2010-1}
S. Tacchi et al., Phys. Rev. B, \textbf{82}, 024401 (2010).

\bibitem{Guslienko}
K. Yu. Guslienko \textsl{et al.}, Phys. Rev. B, \textbf{66}, 132402
(2002).

\end{thebibliography}
\end{document}